\documentclass[preprint]{elsarticle}

\newcommand{\Lnhalf}{L_n^{\left(\frac{1}{2}\right)}}
\newcommand{\bfk}{{\bf k}}
\newcommand{\bfv}{{\bf v}}
\newcommand{\epssqquant}{\left|\epsilon \left(k,\bfk \cdot \bfv + \frac{\hbar k^2}{2 m_e} \right) \right|^2}
\newcommand{\bfvpr}{{\bf v}'}
\newcommand{\betajvsq}{\frac{m_j \beta_j v^2}{2}}
\newcommand{\phiei}{|\phi_{ei}(k)|^2}
\newcommand{\Lminhalf}{L^{\left(-\frac{1}{2}\right)}}
\newcommand{\sspr}{{\sigma \sigma'}}
\newcommand{\Pbar}{\overline{P}}
\newcommand{\Qbar}{\overline{Q}}
\newcommand{\Lhalf}{L^{\left(\frac{1}{2}\right)}}

\journal{Journal of Computational Physics}

\begin{document}

\begin{frontmatter}

\title{Adaptive spectral solution method for the Landau and Lenard-Balescu equations}




\author[LLNL]{Christian R. Scullard \corref{mycorrespondingauthor}}
\ead{scullard1@llnl.gov}
\author[IPAM,PRIN]{Abigail Hickok}
\author[IPAM,UNM]{Justyna O. Sotiris}
\author[IPAM,JH]{Bilyana M. Tzolova}
\author[IPAM,B]{R. Loek Van Heyningen}
\author[LLNL]{Frank R. Graziani}

\cortext[mycorrespondingauthor]{Corresponding author}

\address[LLNL]{Lawrence Livermore National Laboratory}
\address[IPAM]{Institute for Pure and Applied Mathematics, UCLA, Los Angeles CA 90095, USA}
\address[PRIN]{Department of Mathematics, Princeton University, Princeton, NJ 08544, USA}
\address[UNM]{Department of Mathematics and Statistics, University of New Mexico, MSC01 1115, Albuquerque, NM 87131, USA}
\address[JH]{Department of Mathematics, Johns Hopkins University, Baltimore, Maryland 21218, USA}
\address[B]{Department of Mathematics, University of California, Berkeley, CA 94720, USA}

\begin{abstract}
We present an adaptive spectral method for solving the Landau/Fokker-Planck equation for electron-ion systems. The heart of the algorithm is an expansion in Laguerre polynomials, which has several advantages, including automatic conservation of both energy and particles without the need for any special discretization or time-stepping schemes. One drawback is the $O(N^3)$ memory requirement, where $N$ is the number of polynomials used. This can impose an inconvenient limit in cases of practical interest, such as when two particle species have widely separated temperatures. The algorithm we describe here addresses this problem by periodically re-projecting the solution onto a judicious choice of new basis functions that are still Laguerre polynomials but have arguments adapted to the current physical conditions. This results in a reduction in the number of polynomials needed, at the expense of increased solution time. Because the equations are solved with little difficulty, this added time is not of much concern compared to the savings in memory. To demonstrate the algorithm, we solve several relaxation problems that could not be computed with the spectral method without re-projection. Another major advantage of this method is that it can be used for collision operators more complicated than that of the Landau equation, and we demonstrate this here by using it to solve the non-degenerate quantum Lenard-Balescu equation for a hydrogen plasma.
\end{abstract}


\end{frontmatter}


\section{Introduction}
The Landau equation, or its equivalent formulation in terms of the Fokker-Planck equation \cite{Rosenbluth}, is a primary tool in kinetic theory \cite{Bonitz}. In plasma physics, for example, it describes a plasma at conditions for which the Coulomb logarithm dominates the collision integrals, i.e., at weak coupling. Here, the kinetic energy of the particles dominates the potential and small-angle grazing collisions tend to be the most important scattering events. The numerical solution of this system presents many challenges; in particular, great care must be taken when using velocity discretization to ensure conservation of particles and energy \cite{Chang,Epperlein}. Although discretization methods are standard in the field, a spectral method, consisting of an expansion in Laguerre polynomials, was recently used \cite{Scullard} to solve the quantum Lenard-Balescu equation and several variants, including the Landau equation. The motivation for using this technique is that it neatly handles the integration over the dielectric function, but another of its nice properties is that it identically conserves particles and energy without requiring any special method for time stepping. The main drawback is that the method has expensive memory requirements; if one expands the distribution in $N$ polynomials, $N^3$ coefficients must be computed from the equation. If there are $s$ particle species, this increases to $(s^2-s+1)N^3$, so one can quickly find oneself with resolution problems.

Here, we aim to address this difficulty by proposing an adaptive method in which the solution is periodically re-projected onto a new set of basis functions that are better suited to representing the current physical conditions. In this way, we greatly reduce the number of coefficients required, at the expense of increasing the time it takes to solve the actual differential equations. We test the algorithm on electron-ion temperature equilibration, as this problem encapsulates the challenge to the spectral method, namely representing distributions whose temperatures vary widely between species and undergo large variations in time. With the original method, one is limited in how far one can separate the temperatures of the two species; increasing the temperature ratio means needing more polynomials to represent the distribution. Here, we show that this restriction is removed with the adaptive method and we can simulate an arbitrarily large temperature separation. We show also that the algorithm is effective in like-mass cases, where the particles do not remain Maxwellian over the course of the equilibration.

In addition to the Landau equation, our method can be adapted to solve a range of other kinetic equations, and we demonstrate this here by solving the non-degenerate quantum Lenard-Balescu equation. This equation has rarely been attempted numerically \cite{Dolinsky, Scullard} because it requires an integration in frequency and wavenumber over a dielectric function. This dielectric function adds another degree of non-linearity to the system because it contains the distribution function, although this is not the main source of difficulty. The real problem is that under some circumstances, the integrand contains very sharp peaks that are difficult to capture by discretizating either in velocity or over wavenumber and frequency. The spectral method, along with a fairly innocuous approximation in the dielectric function, allows us to circumvent this problem and compute these integrals to good accuracy and quickly enough to be used a numerical solution of the equation. We demonstrate the accuracy of the method by using it to compute the electron-ion equilibration rate in the coupled-mode regime, where missing the sharp peak in the integrand can significantly affect the answer. Our results perfectly match those obtained with sophisticated techniques \cite{Vorberger,Benedict17} that were developed to handle this problem. Of course, we can do more than just compute the rate, which is just a single number, but we can compute entire time evolutions under the Lenard-Balescu equation. We give a few examples in section \ref{sec:num_QLB}. We find, however, that there is not much difference between calculating the distributions with the full Lenard-Balescu collision operator versus using Landau with a well-chosen Coulomb logarithm, even in cases where the temperature separation is large. This is consistent with previous numerical comparisons of the two equations \cite{Scullard,Dolinsky}.

\section{Equation}
To get at the Landau equation, we start with the non-degenerate quantum Lenard-Balescu equation. To use the spectral method, we must calculate the ordinary differential equations satisfied by the expansion coefficients, and this is somehow easier with $\hbar$ non-zero. The Landau equation then appears in the limit $\hbar \rightarrow 0$ and as $\epsilon$, the dielectric function (to be defined below), goes to $1$. The quantum version of the Lenard-Balescu equation features a completely convergent collision operator; a dielectric function provides convergence at large scales, $k \rightarrow 0$, and quantum diffraction ensures it at small scales, $k \rightarrow \infty$. The kinetic equations for the electrons and ions are given by
\begin{eqnarray}
 \frac{\partial f_e}{\partial t} &=& C_{ee}(f_e)+C_{ei}(f_e,f_i)   \label{eq:QLBe} \\ 
 \frac{\partial f_i}{\partial t} &=& C_{ii}(f_i)+C_{ie}(f_e,f_i) . \label{eq:QLBi}
\end{eqnarray}
where
\begin{eqnarray}
& &C_{ei}(f_e,f_i)=-\frac{1}{4 \pi^2 \hbar^2} \int d^3 {\bf v'} \int d^3 {\bf k} \frac{\phiei}{\epssqquant} \cr
&\times&\delta[\bfk \cdot (\bfv-\bfvpr)+\hbar k^2/2 \mu][f_e(\bfv) f_i(\bfvpr)-f_e(\bfv+\hbar \bfk/m_e)f_i(\bfvpr-\hbar \bfk/m_i)],\ \ \  \label{eq:Cei}
\end{eqnarray}
and $\hbar$ is Planck's constant, $\mu$ the reduced mass,
\begin{equation}
 \mu \equiv \frac{m_e m_i}{m_e+m_i},
\end{equation}
and $m_e$ and $m_i$ are the electron and ion masses. We use the Coulomb potential,
\begin{equation}
 \phi_{ie}(k)=\frac{4 \pi Z_i e^2}{k^2} ,
\end{equation}
where $e$ is the electron charge and $Z_i$ is the charge state of species $i$. Note that (\ref{eq:Cei}) defines all the collision integrals in (\ref{eq:QLBe}) and (\ref{eq:QLBi}); to get $C_{ee}$, for example, one just sets $i=e$. In the dielectric function we take the limit $\hbar \rightarrow 0$, which has almost no effect on the solution \cite{Williams,Scullard18} and results in the form
\begin{equation}
 \epsilon_{ie}(X,Y)=1+ \frac{\eta^2_{ie}}{X^2}\left[\frac{\gamma}{Z}w_e \left(\alpha_e^{(ie)}Y \right)+w_i \left(\alpha_i^{(ie)}Y \right) \right], \label{eq:dielectric}
\end{equation}
where the functions $w(x)$ have been derived elsewhere \cite{Scullard} and are given in section \ref{sec:QLB}. As already mentioned, we get the Landau equation by taking $\hbar \rightarrow 0$ everywhere in the equation and setting $\epsilon=1$. This renders the $k$-integral divergent at both of its limits; these must then be cut off, resulting in a Coulomb logarithm. We now discuss our solution method and its implementation.

\section{Method}
\subsection{Expansion}
We use the expansion in Laguerre polynomials used in Ref. \cite{Scullard} to solve the quantum Lenard-Balescu equation. For the species $j$,
\begin{equation}
 f_j(v,t)=f_j^{\mathrm{eq}}(v) \sum_{n=0}^{N} A^j_n(t) \Lnhalf \left(\frac{\beta_j m_j v^2}{2} \right) \label{eq:expansion}
\end{equation}
where $f_j^{\mathrm{eq}}(v)$ is the Maxwell distribution at the temperature $\beta_j$,
\begin{equation}
 f_j^{\mathrm{eq}}(v) \equiv n_j \left(\frac{m_j \beta_j}{2 \pi}\right)^{3/2} \exp \left(-\betajvsq \right),
\end{equation}
and $L_n^{(1/2)}(x)$ is the $n^{\mathrm{th}}$ Laguerre polynomial of order $1/2$. We stress that the temperature $\beta_j$ is just an arbitrary expansion parameter and need not be related to any real property of the system. This parameter plays a central role in the adaptive method.

The equations for the $A_m(t)$ are found by multiplying both sides of the equation by
\begin{equation}
 L_m^{\left(\frac{1}{2}\right)}\left(\betajvsq\right)
\end{equation}
and integrating over $d^3 \bfv$. The resulting coupled ordinary differential equations are then of the form
\begin{eqnarray}
 \frac{d A_n^e}{dt}&=& \sum_{n'n''} C^{(ee)}_{nn'n''} A^e_{n'}(t)A^e_{n''}(t)+\sum_{n'n''} C^{(ei)}_{nn'n''} A^e_{n'}(t)A^i_{n''}(t) \label{eq:Ae} \\ 
 \frac{d A_n^i}{dt}&=& \sum_{n'n''} C^{(ii)}_{nn'n''} A^i_{n'}(t)A^i_{n''}(t)+\sum_{n'n''} C^{(ie)}_{nn'n''} A^i_{n'}(t)A^e_{n''}(t) \label{eq:Ai}
\end{eqnarray}
where, for the Landau equation, the $C^{(jk)}_{nn'n''}$ are constants proportional to $\log \Lambda$. These will be given in section \ref{sec:Landau_coeff} and those of the quantum Lenard-Balescu equation in section \ref{sec:QLB}.

\subsection{Conservation}
Particles will be conserved exactly provided
\begin{equation}
 A^j_0(t)=1 \label{eq:particle_cons}
\end{equation}
is satisfied throughout the calculation (this simple relation is the reason for the choice of $1/2$ for the order of the Laguerre polynomials). Equations (\ref{eq:Ae}) and (\ref{eq:Ai}) have the property that
\begin{equation}
 \frac{d A_0^e}{dt}=\frac{d A_0^i}{dt}=0
\end{equation}
identically, so that (\ref{eq:particle_cons}) is trivially satisfied if it holds initially. As for conservation of energy, its expression in terms of the $A_n$ is
\begin{equation}
 A_1^e = -\frac{n_i \beta_e}{n_e \beta_i}A_1^i , \label{eq:energy_cons}
\end{equation}
which can easily be derived by computing the kinetic energy integrals for each species and demanding their sum be constant. The equations 
(\ref{eq:Ae}) and (\ref{eq:Ai}) have the property that
\begin{equation}
 \frac{dA_1^e}{dt}=-\frac{n_i \beta_e}{n_e \beta_i}\frac{dA_1^i}{dt}
\end{equation}
which arises from the identities
\begin{eqnarray}
 C^{(ee)}_{1n'n''}&=&-C^{(ee)}_{1n''n'} \\
 C^{(ii)}_{1n'n''}&=&-C^{(ii)}_{1n''n'} \\
 C^{(ei)}_{1n'n''}&=&-\frac{n_i \beta_e}{n_e \beta_i}C^{(ie)}_{1n''n'},
\end{eqnarray}
so that the method guarantees conservation of energy.

\subsection{Adaptive temperature}
The parameters $\beta_j$ are (effective) inverse temperatures and can be chosen arbitrarily; it is this freedom that we exploit in the adaptive method. Clearly, if the distribution we wish to resolve is in fact Maxwellian at the temperature $\beta$, then choosing $\beta_j=\beta$ means that the expansion is simply $A_n^j=\delta_{n0}$ and we need only one coefficient to represent the distribution. Choosing a different $\beta_j$ would require more coefficients, with the number depending on how far $\beta$ is from $\beta_j$. Although this is a somewhat trivial example, it is generally the case that some choices of $\beta_j$ are better than others from the standpoint of minimizing $N$. Moreover, a good choice at one point in the solution may be very poor at a later time. For example, if we consider electron-ion temperature equilibration, the species are roughly Maxwellian over their entire evolution with only their temperatures changing in time. For species $j$, if we set $\beta_j$ to its initial temperature then we can easily resolve the initial condition, but as time goes on the actual temperature of the distribution becomes very different from this $\beta_j$. If the initial temperature separation of the two species was large enough, we will run out of resolution with $N$ coefficients. The obvious remedy to this problem is to recalculate, at various intervals, the expansion coefficients using the current temperature, with the result that most will return to zero. Although this only sounds useful when the distributions remain approximately Maxwellian, the latter is often the case in systems of particles with large mass disparities such as electrons and ions. Furthermore, using an effective temperature in (\ref{eq:expansion}) that corresponds to the kinetic energy of the distribution is usually the most efficient choice even when particles are not Maxwellian \cite{Scullard}.

If we are given a set of coefficients $A_i$, for any species, corresponding to the temperature parameter $\beta$, and we wish to find the coefficients $A_i'$ corresponding to $\beta'$, the formula is
\begin{equation}
 A_n'=n! \sum_{j=0}^n (-1)^{n+j} A_j \frac{\left(\frac{\beta'}{\beta}\right)^j \left(\frac{\beta'}{\beta}-1\right)^{n-j}}{j!(n-j)!} . \label{eq:Aprime}
\end{equation}
The derivation is given in \ref{sec:reprojection}. Note that this transformation has no effect on particle or energy conservation as
\begin{equation}
 A'_0=A_0=1
\end{equation}
and if (\ref{eq:energy_cons}) is satisfied, then conservation of kinetic energy and (\ref{eq:Aprime}) guarantee that (\ref{eq:energy_cons}) is also satisfied in the primed variables.

The next issue to consider is when to actually perform these reprojections. The most obvious thing to do is to find the $\beta'$ associated with the current kinetic energy and check its ratio with the $\beta$ used in the expansion, $\beta'/\beta$. When its deviation from unity is larger than some predefined tolerance, we trigger a reprojection. Computing this $\beta'$ is a simple matter, and we derive the formula in \ref{sec:betaprime},
\begin{equation}
 \beta' = \frac{\beta}{1-A_1(\beta)}. \label{eq:betaprime}
\end{equation}
For each species $\sigma$, we define the quantity
\begin{equation}
 \Delta_\sigma \equiv \left|\frac{\beta_{\sigma}'}{\beta_{\sigma}}-1 \right|
\end{equation}
and if $\Delta_\sigma > c$ we reproject. The smaller we make $c$, the more frequent will be the reprojections, which will generally lower the number of polynomials needed. Of course, there are other alternatives. One is to look at the size of the highest-order coefficients and to reproject when they appear to be getting too large. Another is simply to reproject at regular intervals. We will give examples of all three schemes in action below.

\subsection{Equation coefficients} \label{sec:coeff}
Deriving the coefficients $C_{nn'n''}$ is a straightforward but tedious task. We do not outline the steps here, but refer the reader to Ref. \cite{Scullard}, where the calculation was done in detail for the one-component, single-temperature case. The generalization to multiple species with different temperatures is fairly obvious and requires nothing conceptually different from what was presented there. Although the collision operators defined in (\ref{eq:Cei}) are 6-fold integrals, and to get the coefficients we must integrate these further over $d^3 {\bf v}$, the final 9-fold integrals can be reduced by various identities to
\begin{eqnarray}
& &C^{(\sigma \sigma')n}_{lk}=-\frac{n_e \sqrt{2 \pi} b_2^{(\sigma \sigma')}}{2 \sqrt{b_1^{(\sigma \sigma')}}} (\beta_\sigma m_\sigma)^{1/2}(\beta_{\sigma'} m_{\sigma'})^{1/2} e^4 Z_\sigma^2 Z_{\sigma'}^2 \frac{n!}{\Gamma(n+3/2)} \cr
&\times& \int_{-\infty}^\infty \int_0^\infty \frac{1}{X^3} \frac{e^{-Y^2-X^2}}{|\epsilon_{\sigma \sigma'}(X,Y)|^2} q^{(\sigma \sigma')n}_{lk}(X,Y) dXdY \label{eq:CnlkXY}
\end{eqnarray}
where
\begin{eqnarray}
& & q^{(ie)n}_{lk}(X,Y) \equiv \sum_{j=0}^{\min(l,n)} \Lminhalf_{n-j} \left(\overline{Y}^2_- \right) \cr
&\times& \left[e^{-\hbar \omega \Delta \beta/2} \Lminhalf_{l-j} \left(\overline{Y}^2_- \right) \Lminhalf_k \left(Y^2_+ \right) - e^{\hbar \omega \Delta \beta/2} \Lminhalf_{l-j} \left(\overline{Y}^2_+ \right) \Lminhalf_k \left(Y^2_- \right) \right],\ \ \ \ \ \  \label{eq:qnlk}
\end{eqnarray}
$\Delta \beta_{ie} \equiv \beta_e-\beta_i$, and the various dimensionless variables are
\begin{eqnarray}
 Y^2 &\equiv& \frac{b_1^{(\sigma \sigma')}}{2} \frac{\omega^2}{k^2} \\
 X^2 &\equiv& \frac{\hbar^2 b_2^{(\sigma \sigma')}}{8} k^2 \\
 \overline{Y}_{\pm}&=&\overline{P}_{ie} Y \pm \overline{Q}_{ie} X \\
 Y_{\pm}&=&P_{ie} Y \pm Q_{ie} X 
\end{eqnarray}
where
\begin{eqnarray}
 b_1^{(\sigma \sigma')} &\equiv& m_i \beta_i+m_e \beta_e \\
 b_2^{(\sigma \sigma')} &\equiv& \frac{\beta_i}{m_i}+\frac{\beta_e}{m_e} \\
 \overline{P}_{ie}&=& \sqrt{\frac{\beta_i m_i}{\beta_i m_i + \beta_e m_e}} \\
 \overline{Q}_{ie}&=& \sqrt{\frac{\beta_i m_e}{\beta_i m_e + \beta_e m_i}} \\
 P_{ie}&=& \sqrt{\frac{\beta_e m_e}{\beta_i m_i + \beta_e m_e}} \\
 Q_{ie}&=& \sqrt{\frac{\beta_e m_i}{\beta_e m_i + \beta_i m_e}}\ .
\end{eqnarray}
For the dielectric function we use the classical approximation; that is, we neglect quantum diffraction \cite{Scullard}. The main benefit of this is that the dielectric function takes the comparatively simple form (\ref{eq:dielectric}). That is, the variables $X$ and $Y$ separate, facilitating analytical evaluation of the $X$-integral without losing any important physical content. The constant $\gamma$ is the temperature ratio 
\begin{equation}
 \gamma \equiv \frac{\beta_e}{\beta_i}
\end{equation}
and the $\eta$ are given by
\begin{equation}
 \eta_{ie} \equiv \frac{\lambda_Q^{(ie)}}{\lambda_D} \label{eq:eta}
\end{equation}
with
\begin{eqnarray}
 \lambda_Q^{(ie)} &\equiv& \frac{\hbar}{2 \sqrt{2}}\sqrt{\left(\frac{\beta_e}{m_e}+\frac{\beta_i}{m_i} \right)} \\
 \lambda_D &\equiv& \frac{1}{4 \pi e^2 Z n_e \beta_i}.
\end{eqnarray}
The complex function, $w(Y)$, is given by \cite{Scullard},
\begin{eqnarray}
 \mathrm{Re} [w_e(x)] &=& \sum_{l=0}^{\infty} A_l^e M \left(l+1,\frac{1}{2};-\frac{x^2}{2} \right)\\
 \mathrm{Im} [w_e(x)] &=& \sqrt{\frac{\pi}{2}}x e^{-x^2/2} \sum_{l=0}^{\infty} A_l^e \Lhalf_l\left(\frac{x^2}{2}\right)
\end{eqnarray}
and the constants $\alpha$ are
\begin{eqnarray}
 \alpha_e^{(ii)}&=&\sqrt{\frac{\beta_e m_e}{\beta_i m_i}} \\
 \alpha_e^{(ie)}&=& \alpha_e^{(ei)} = \sqrt{\frac{2 \beta_e m_e}{\beta_i m_i+\beta_e m_e}} \\
 \alpha_i^{(ee)}&=&\sqrt{\frac{\beta_i m_i}{\beta_e m_e}} \\
 \alpha_i^{(ie)}&=& \alpha_i^{(ei)} = \sqrt{\frac{2 \beta_i m_i}{\beta_i m_i+\beta_e m_e}} \\
 \alpha_i^{(ii)}&=& \alpha_e^{(ee)} =1 .
\end{eqnarray}

\subsection{Landau equation} \label{sec:Landau_coeff}
So far, we have kept $\hbar$ in the calculation. Now, we can set it to zero by making $X=0$ except in $dX/X$ where the $\hbar$ will cancel. If we also set $\epsilon(X,Y)=1$, the result is the Landau equation. Only the terms of $q(X,Y)$ that are even in $Y$ survive the integration in (\ref{eq:CnlkXY}) and so we write
\begin{equation}
 P(X,Y) \equiv \frac{1}{2}\left[q(X,Y)+q(X,-Y)\right]. \label{eq:Pdef}
\end{equation}
Now, we separate this function into a piece proportional to $X^2$ and one containing all higher-order terms in $X$,
\begin{equation}
 P(X,Y) = X^2 Y^2 G(Y) + R(X,Y) \label{eq:p_expansion}
\end{equation}
where $G(Y)$ is a finite polynomial in $Y$ and $R(X,Y)$ is $\mathrm{O}(X^4)$. Clearly, the terms $X^4$ and higher vanish in the integrand in (\ref{eq:qnlk}) when $X$ is set to zero, but the $X^2$ term gives us
\begin{equation}
 \int \frac{dX}{X} \rightarrow \ln \Lambda_{ie},
\end{equation}
i.e., the Coulomb logarithm. Thus we have
\begin{eqnarray}
& &C^{(\sigma \sigma')n}_{lk}=-\frac{n_e \sqrt{2 \pi} b_2^{(\sigma \sigma')}}{2 \sqrt{b_1^{(\sigma \sigma')}}} (\beta_i m_i)^{1/2}(\beta_e m_e)^{1/2} e^4 Z_\sigma^2 Z_{\sigma'}^2  S_{nn'n''} \ln \Lambda_{\sigma \sigma'} \label{eq:ClnLambda}
\end{eqnarray}
where
\begin{equation}
 S_{nn'n''} \equiv \frac{n!}{\Gamma(n+3/2)}\int_{-\infty}^\infty e^{-Y^2} Y^2 G_{nn'n''}(Y) dY . \label{eq:Sn}
\end{equation}
The polynomials $G(Y)$ are given by
\begin{eqnarray}
& & G^{(\sspr)n}_{lk}(Y) =\frac{1}{2} \left. \frac{\partial^2}{\partial X^2}q^{(\sspr)n}_{lk}(X,Y) \right|_{X=0} \cr
 &=&8 \Pbar_\sspr \Qbar_\sspr \sum_{j=0}^{\min(l,n)} \Lhalf_{n-j-1}(\Pbar_\sspr^2 Y^2) \left\{ -P_\sspr Q_\sspr \Lhalf_{k-1}(P_\sspr^2 Y^2) \Lminhalf_{l-j} (\Pbar_\sspr^2 Y^2) \right. \cr
 & & + \left. \Lminhalf_k(P^2_\sspr Y^2) \left[ \Pbar_\sspr \Qbar_\sspr \Lhalf_{l-j-1}(\Pbar^2_\sspr Y^2) - \frac{\Delta \beta_{\sspr}}{\sqrt{b_1^{(\sigma \sigma')} b_2^{(\sigma \sigma')}}} \Lminhalf_{l-j}(\Pbar_\sspr^2 Y^2) \right] \right\}. \label{eq:Gn}
\end{eqnarray}
where
\begin{equation}
 \beta_{\sspr} \equiv \beta_{\sigma'}-\beta_{\sigma} .
\end{equation}
The bulk of the work in the adaptive spectral method is in computing the coefficients $S_{nn'n''}$; the resulting ordinary differential equations are quicky solved with a standard Runge-Kutta method. To determine the coefficients, we use Gaussian quadrature to compute their defining integral, equation (\ref{eq:Sn}). To do this, we first observe that all $G(Y)$ are even, as can easily be deduced by glancing at (\ref{eq:Gn}). This suggests the substitution $x=Y^2$ and we then have
\begin{equation}
  S_{nn'n''} \equiv \frac{n!}{\Gamma(n+3/2)}\int_0^\infty x^{1/2} e^{-x} G_{nn'n''}(\sqrt{x}) dx . \label{eq:Snx}
\end{equation}
Clearly, the quadrature scheme should be based on associated Laguerre polynomials of order $1/2$ as the weight function is $x^{1/2}e^{-x}$. To determine the number of points, $N_p$, we consider that Gaussian quadrature is exact for integrands consisting of polynomials of degree up to $2N_p-1$ that multiply the weight. From (\ref{eq:Gn}) it can easily be inferred that the maximum degree of $G_{nn'n''}(\sqrt{x})$ is around $3N$ (i.e., when $n=n'=n''=N$), although in practice it is usually slightly less. We therefore have
\begin{equation}
 N_p \approx \frac{3N}{2} .
\end{equation}

Once $N_p$ has been chosen, we need only the weights, $W_j$, and quadrature points, $x_j$. The latter are given by the roots of the $N_p^{\mathrm{th}}$ Laguerre polynomial of order $1/2$,
\begin{equation}
 \Lhalf_{N_p}(x_j)=0,
\end{equation}
and the weights are
\begin{equation}
 W_j=\frac{x_j \Gamma(N_p+1/2)}{N_p! (N_p+1/2) \left[L_{N_p-1}^{(1/2)}(x_j)\right]^2} \ . \label{eq:weights}
\end{equation}
The integral determining $S_{nn'n''}$ is then given by
\begin{equation}
 \int_0^\infty x^{1/2} e^{-x} G_{nn'n''}(\sqrt{x}) dx = \sum_{j=1}^{N_p} W_j G_{nn'n''}(\sqrt{x_j}), \label{eq:quad}
\end{equation}
provided $N_p$ is large enough, otherwise (\ref{eq:quad}) is an approximation. The final ingredient needed is an efficient way to determine $G_{nn'n''}(\sqrt{x})$ at the quadrature points. For this, we need to determine $\Lhalf_n(\xi x_j)$ and $\Lminhalf_n(\xi x_j)$ for all $n$ up to $N$, all $j$ up to $N_p$, and a variety of $\xi$. The best way to do this is to use the formulas
\begin{eqnarray}
 L_0^{\left(\alpha \right)}(x)&=&1 \\
 L_1^{\left(\alpha \right)}(x)&=&1 + \alpha -x 
\end{eqnarray}
and the recurrence relation
\begin{equation}
 L_n^{(\alpha)}(x)=\left(2+\frac{\alpha-1-x}{n}\right) L_{n-1}^{(\alpha)}(x) - \left(1 + \frac{\alpha-1}{n}\right) L_{n-2}^{(\alpha)}(x) \  \label{eq:Lrecur}
\end{equation}
for each $\xi x_j$. The formula (\ref{eq:Lrecur}) is numerically stable so there is no problem going to large $n$. Equation (\ref{eq:quad}) must be evaluated every time there is a re-projection because of the temperature dependences in various parts of (\ref{eq:Gn}). 

We now have what we need to compute the coefficients of the ordinary differential equations for the Landau system. To solve these equations, we use the fourth-order Runge-Kutta method as implemented in the Boost C++ library \cite{Schling}. Numerical results are given in Sec. \ref{sec:num_Landau}.

\subsection{Quantum Lenard-Balescu equation} \label{sec:QLB}
To solve the QLB equation, we need a way to compute efficiently the integral over the dielectric function in Eq. (\ref{eq:CnlkXY}). The strategy is to exploit the separation between the variables $X$ and $Y$ in the Eq. (\ref{eq:dielectric}) and to do the $X$-integral analytically. In fact, we will be interested only in the order-unity term and the one proportional to $\ln \eta$. At weak coupling, these are by far the dominant terms in the collision operator.

Unlike in the Landau equation, we do not take the limit $X \rightarrow 0$. As demonstrated in Ref. \cite{Scullard}, the terms $\mathrm{O}(X^4)$ and higher in the expansion (\ref{eq:p_expansion}) do not require the dielectric function for their integrals to converge. This means that we may drop the dielectric function from those integrals as it will contribute terms higher order in $\eta$ than are of interest to us here. However, the integral over $R(X,Y)$ does contribute an order-unity term, which we call $B_{nn'n''}^{(\sigma \sigma')}$ and is defined by
\begin{equation}
 B_{nn'n''}^{(\sigma \sigma')} \equiv \frac{n!}{\Gamma(n+3/2)}\int_0^{\infty} dX \frac{e^{-X^2}}{X^3}\int_{-\infty}^{\infty} dY e^{-Y^2} R^{(\sigma \sigma')}_{nn'n''}(X,Y) .
\end{equation}
where $R^{(\sigma \sigma')}_{nn'n''}(X,Y)$ can be calculated from (\ref{eq:p_expansion}). To evaluate this, we use Gauss-Laguerre quadrature with a small modification from what was done in the previous section. Making the substitutions $y=X^2$ and $x=Y^2$,
\begin{equation}
 B_{nn'n''}^{(\sigma \sigma')} \equiv \frac{n!}{\Gamma(n+3/2)}\int_0^{\infty} dy \frac{e^{-y}}{y^2}\int_{0}^{\infty} dx x^{1/2} e^{-x} \frac{R^{(\sigma \sigma')}_{nn'n''}(\sqrt{y},\sqrt{x})}{x} . \label{eq:Bxy}
\end{equation}
Because the lowest order term in $R(\sqrt{y},\sqrt{x})$ is $y^2$, the $1/y^2$ in Eq. (\ref{eq:Bxy}) will cancel, leaving the weight $e^{-y}$. The $1/x$ will be similarly cancelled and thus we see that the best choice here is to use the order $0$ Laguerre polynomials for the $y$-integral and order $1/2$ for $x$,
\begin{equation}
 B_{nn'n''}^{(\sigma \sigma')} \approx \frac{1}{2} \sum_i^{N_p} \sum_j^{N_p} W_i W_j^{(0)} \frac{1}{x_i y_j^2} R^{(\sigma \sigma')}_{nn'n''}(\sqrt{y_j},\sqrt{x_i})
\end{equation}
where $x_i$ are given by Eq. (\ref{eq:weights}), $y_i$ are the zeros of the order zero (or unassociated) Laguerre polynomials,
\begin{equation}
 L_{N_p}(y_j)=0 ,
\end{equation}
and the weights $W_j^{(0)}$ are
\begin{equation}
 W_j^{(0)} = \frac{y_j}{N_p^2 \left[L_{N_p-1}(y_j)\right]^2} \ .
\end{equation}
The double integration makes $B_{nn'n''}^{(\sigma \sigma')}$ comparatively expensive to calculate. However, as we will show in Sec. \ref{sec:num_QLB}, these coefficients tend to have an insignificant impact on the calculation, at least for the relaxation problems we consider.

The piece of the integral that does require the dielectric function for convergence we write as
\begin{equation}
 I_{nn'n''}^{(\sigma \sigma')} \equiv \int_0^\infty dX \frac{e^{-X^2}}{X}\int_{-\infty}^{\infty} dY \frac{e^{-Y^2}}{|\epsilon_{\sigma \sigma'}(X,Y)|^2}Y^2 G_{nn'n''}^{(\sigma \sigma')}(Y) .
\end{equation}
The expansion of such integrals in $\eta$ was given in Appendix C of Ref. \cite{Scullard} and is easily adapted to the multi-species case,
\begin{equation}
 I_{nn'n''}^{(\sigma \sigma')} \approx -\frac{\Gamma(n+3/2)}{n!} \left[S_{nn'n''}^{(\sigma \sigma')} \left(\frac{\gamma_E}{2}+\ln \eta_{\sigma \sigma'} \right) +F_{nn'n''}^{(\sigma \sigma')} \right]
\end{equation}
where
\begin{equation}
 F^{(\sigma \sigma')}_{nn'n''} \equiv \frac{n!}{\Gamma(n+3/2)}\frac{1}{2} \int_{-\infty}^{\infty} dY e^{-Y^2} Y^2 G_{nn'n''}^{(\sigma \sigma')}(Y)F^{(\sigma \sigma')}(Y)
\end{equation}
and
\begin{eqnarray}
 F^{(\sigma \sigma')}(Y) &\equiv& \frac{1}{2} \ln  \left( \left[w_R^{(\sigma \sigma')}(Y)\right]^2 + \left[w_I^{(\sigma \sigma')}(Y)\right]^2 \right) \cr
 & &+\frac{w_R^{(\sigma \sigma')}(Y)}{w_I^{(\sigma \sigma')}(Y)} \arctan\left[w_R^{(\sigma \sigma')}(Y),w_I^{(\sigma \sigma')}(Y) \right] ,
\end{eqnarray}
$\gamma_E \approx 0.5772156649...$ is the Euler constant, and $w_R^{(\sigma \sigma')}(Y)$ and $w_I^{(\sigma \sigma')}(Y)$ are given in (\ref{eq:dielectric}). The coefficients $S_{nn'n''}^{(\sigma \sigma')}$ are the same as those defined in Sec. \ref{sec:Landau_coeff}.

We know of no simple closed form for the integrals $F^{(\sigma \sigma')}_{nn'n''}$, but despite appearances they are fairly simple to evaluate numerically. We can once again use the Gauss-Laguerre scheme used to compute $S_{nn'n''}^{(\sigma \sigma')}$, so that
\begin{eqnarray}
 \int_{-\infty}^{\infty}dY e^{-Y^2} Y^2 G_{nn'n''}^{(\sigma \sigma')}(Y)F^{(\sigma \sigma')}(Y)&=&\int_0^{\infty}dx x^{1/2} e^{-x} G_{nn'n''}^{(\sigma \sigma')}(\sqrt{x})F^{(\sigma \sigma')}(\sqrt{x}) \cr
 &\approx& \sum_{i=1}^{N_p}W_i G_{nn'n''}^{(\sigma \sigma')}(\sqrt{x_i})F^{(\sigma \sigma')}(\sqrt{x_i})
\end{eqnarray}
with the weights given in (\ref{eq:weights}). Evaluating $F^{(\sigma \sigma')}(\sqrt{x_i})$ is fairly simple; although it contains hypergeometric functions, these need only be evaluated during a reprojection and their values may be stored for subsequent time steps. The GNU Scientific Library \cite{GSL} contains functions that quickly and accurately perform these evaluations. The Laguerre polynomials that appear in $F^{(\sigma \sigma')}(\sqrt{x_i})$ can be evaluated using the recurrence relation given in Sec. \ref{sec:Landau_coeff}.

Putting the various pieces together,
\begin{eqnarray}
C^{(\sigma \sigma')}_{nn'n''}=C_0^{(\sigma \sigma')}\left[S^{(\sigma \sigma')}_{nn'n''}\left(\frac{\gamma_E}{2}+\ln \eta_{\sigma \sigma'} \right)-B^{(\sigma \sigma')}_{nn'n''}+F^{(\sigma \sigma')}_{nn'n''}(\{A^{\sigma},A^{\sigma'}\}) \right] \label{eq:Ccalc}
\end{eqnarray}
where
\begin{equation}
 C_0^{(\sigma \sigma')} \equiv \frac{n_{\sigma'} \sqrt{2 \pi} b^{(\sigma \sigma')}_2}{2 \sqrt{b^{(\sigma \sigma')}_1}} (\beta_\sigma m_\sigma)^{1/2}(\beta_{\sigma'} m_{\sigma'})^{1/2} e^4 Z_\sigma^2 Z_{\sigma'}^2 \ .
\end{equation}
\section{Numerical results}
\subsection{Landau Equation} \label{sec:num_Landau}
\subsubsection{Electron-proton system}
\begin{figure}
  \includegraphics[width=4in]{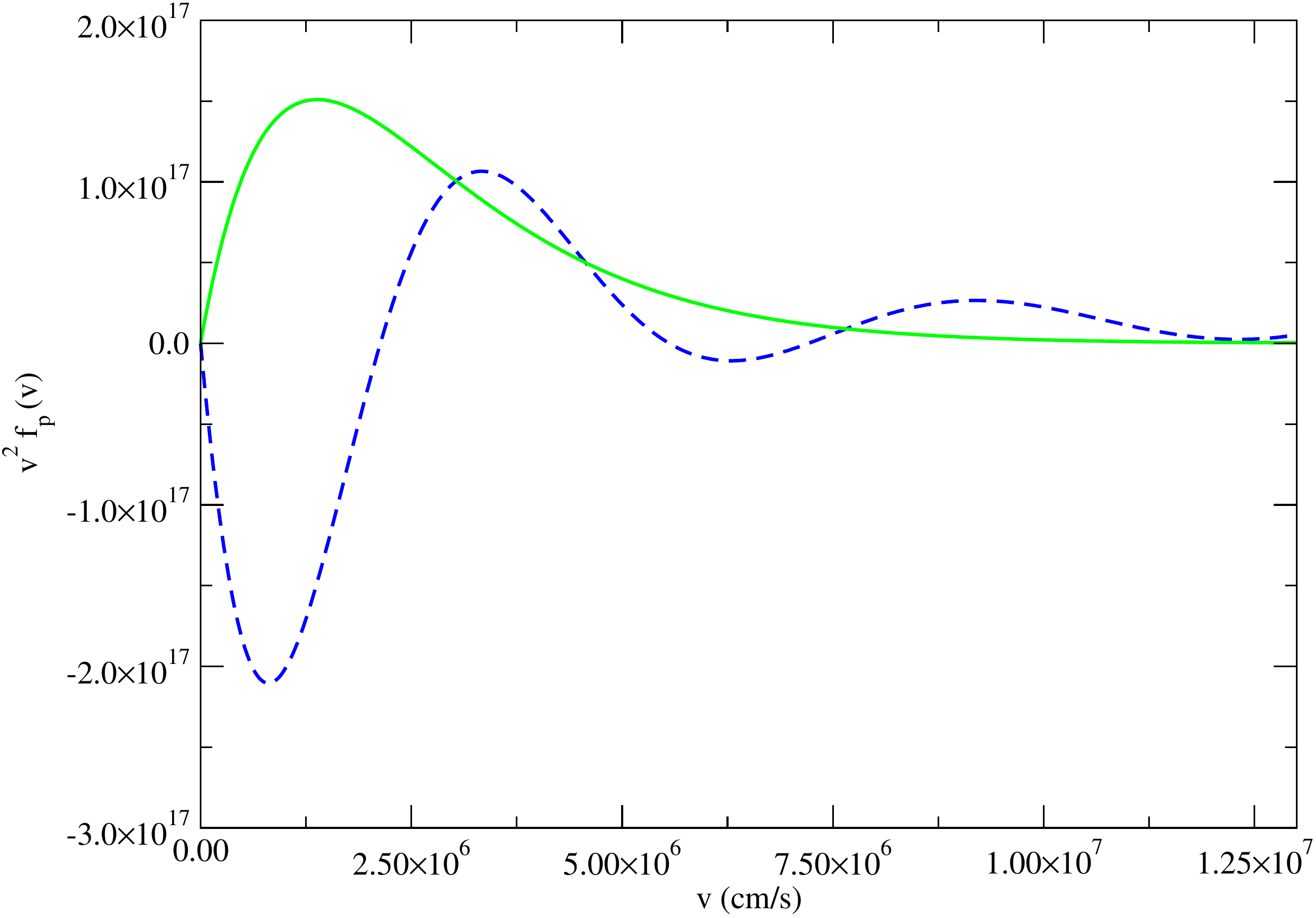}
 \caption{Evolution of the proton distribution for hydrogen with initial electron temperature $T_e = 10 \ \mathrm{keV}$, proton temperature $T_i = 10 \ \mathrm{eV}$, and density $n_e=n_p=10^{25} \ \mathrm{cm}^{-3}$ and $n_{\mathrm{max}}=5$. Solid green: initial distribution; dotted blue; distribution at $t=1.0 \times 10^{-14}\ \mathrm{s}$. We clearly do not have enough resolution to complete this calculation without reprojecting.}
 \label{fig:proton_underres}
\end{figure}

To begin, we demonstrate the resolution problems that can arise when we have too few polynomials and do not use reprojection. We consider a hydrogen system at density $n_e=n_p=10^{25}\ \mathrm{cm}^{-3}$ and with initial electron and proton temperatures $T_e=10\ \mathrm{keV}$ and $T_p=10\ \mathrm{eV}$. This is a highly unrealistic temperature separation, but it serves to illustrate the numerical method. In Fig. \ref{fig:proton_underres}, we plot the proton distribution at the initial time and at $t=1 \times 10^{-14} \ \mathrm{s}$ using six polynomials ($n_{\mathrm{max}}=5$). We can see that as the solution progresses, and the effective temperature becomes more widely separated from the expansion temperature, we become unable to represent the distribution with only six polynomials.

\begin{figure}
  \includegraphics[width=5in]{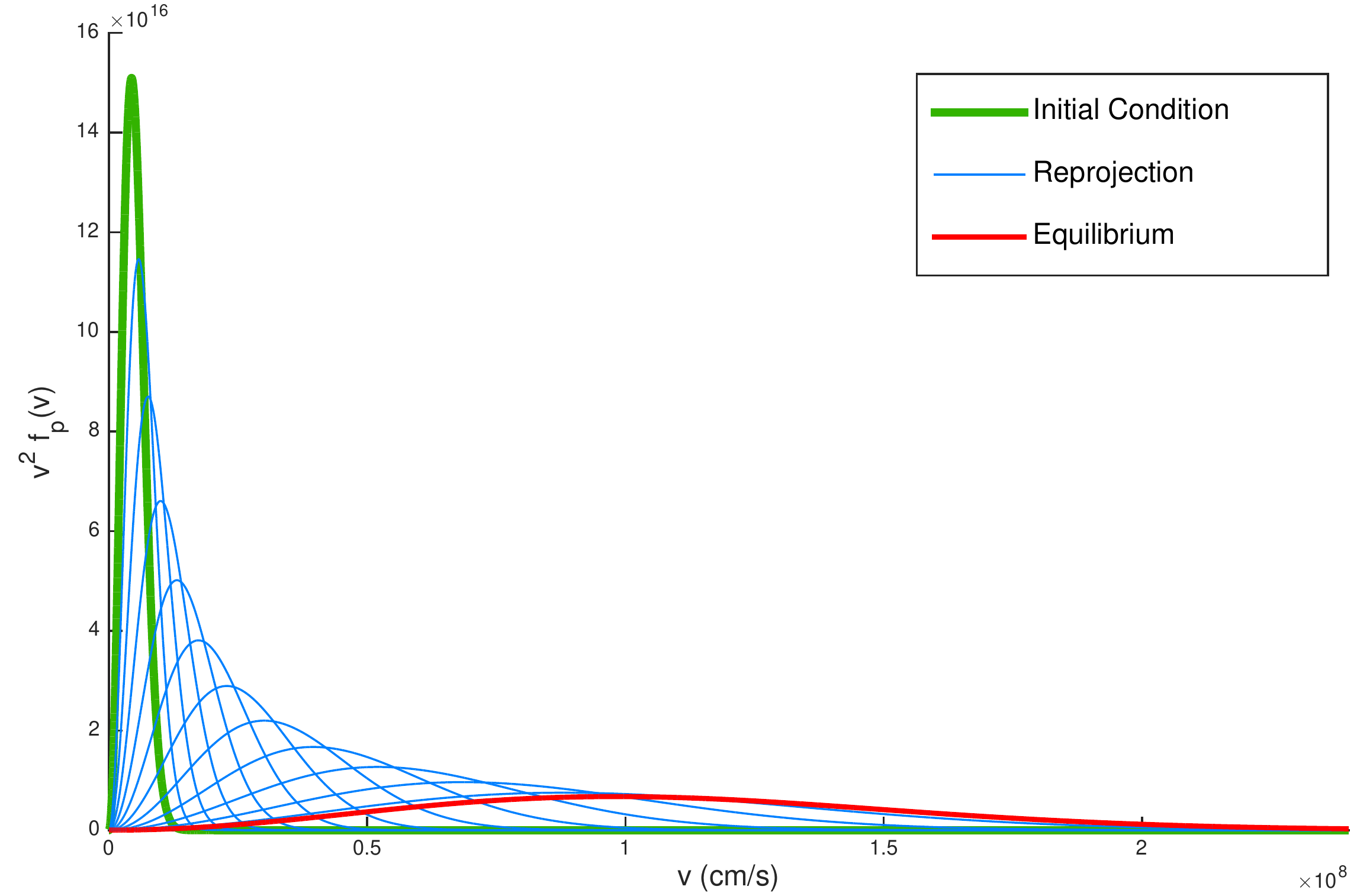}
 \caption{Evolution of the proton distribution for hydrogen with initial electron temperature $T_e = 10 \ \mathrm{keV}$, proton temperature $T_i = 10 \ \mathrm{eV}$, and density $n_e=n_p=10^{25} \ \mathrm{cm}^{-3}$. The distribution is plotted in blue at every second reprojection.}
 \label{fig:Landau_p_reprojection}
\end{figure}

\begin{figure}
  \includegraphics[width=5in]{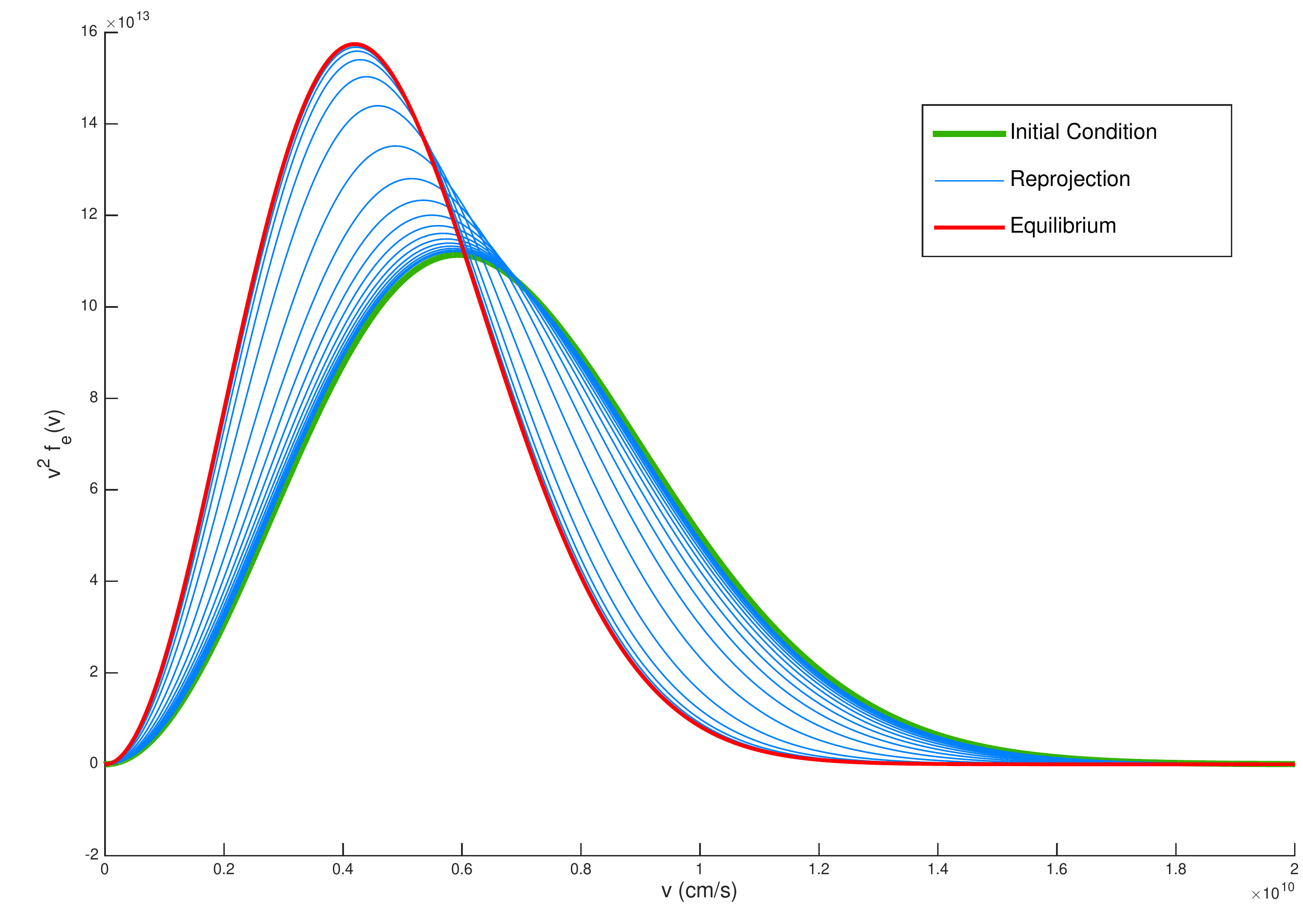}
 \caption{Evolution of the electron distribution for hydrogen with initial electron temperature $T_e = 10 \ \mathrm{keV}$, proton temperature $T_i = 10 \ \mathrm{eV}$, and density $n_e=n_p=10^{25} \ \mathrm{cm}^{-3}$. The distribution is plotted in blue at every second reprojection.}
 \label{fig:Landau_e_reprojection}
\end{figure}

Now we perform the same calculation using a scheme in which a reprojection is triggered when $|A^j_4|>10^{-2}$ for either species $j$. In Fig. \ref{fig:Landau_p_reprojection} we have plotted the proton distribution at the initial time and at every second reprojection. The electron distribution is shown in Fig. \ref{fig:Landau_e_reprojection}. Clearly, we have solved the resolution issue as the calculation now proceeds all the way to equilibrium without any problems. 

\subsubsection{Electron-electron system}
\begin{figure}
  \includegraphics[width=5in]{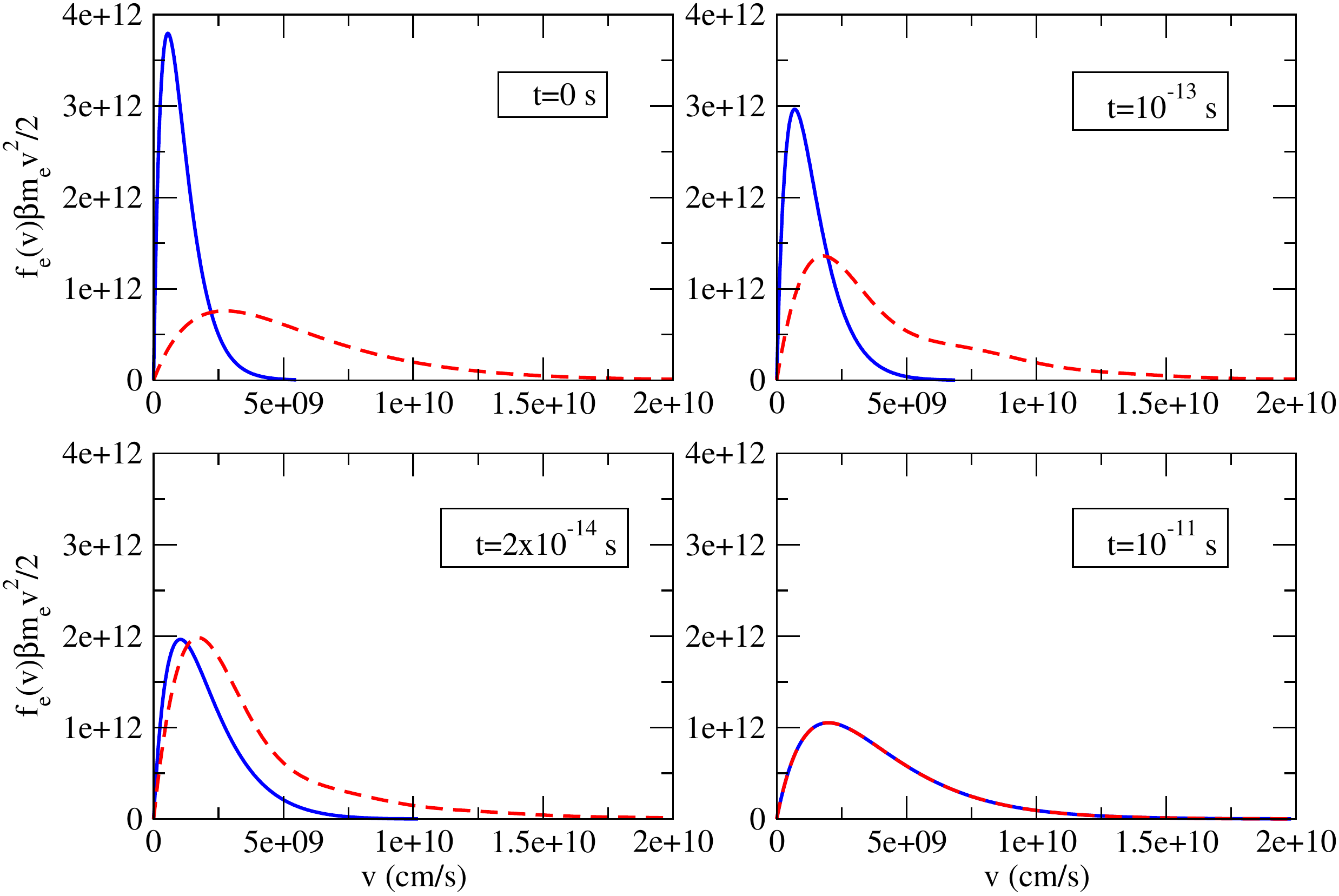}
 \caption{Evolution of an electron-electron system at $n_e=2 \times 10^{23} \ \mathrm{cm}^{-3}$. The system is divided in half into two Maxwellians, one at $T_1=10^7K$ and the other at $T=2.5 \times 10^8K$. The hotter particles are the dotted red lines and the colder the solid blue. The real single-particle distribution is the sum of these two.}
 \label{fig:electron_equilibration}
\end{figure}

In the disparate-mass system, the distributions are basically Maxwellians throughout their whole evolution. The adaptive scheme is clearly ideal for such problems, but we demonstrate here that a wide mass separation is not essential for the method to be useful. We consider electrons at a density $n_e=2 \times 10^{23}\ \mathrm{cm}^{-3}$ where half the particles start in a Maxwell distribution with temperature $T_1=10^7 K$ and the other half have temperature $T_2=2.5 \times 10^8K$, a separation by a factor of 25. In this case, the system will certainly not maintain Maxwellian profiles over the course of the equilibration and we must thus ensure that we have enough polynomials to capture the entire evolution. We take $n_{\mathrm{max}}=20$ for this calculation, which is not enough to complete the calculation without reprojections. We use a scheme of reprojecting at regular intervals, which, as a general rule, should probably be used for cases in which the species do not remain roughly Maxwellian. In fact, for this case we reproject on every time step. This is not really necessary, but running this way minimizes the number of polynomials needed. In practice, one should find a balance between the number of polynomials used and the number of time steps between reprojections. The result is shown in Figure \ref{fig:electron_equilibration}, where various times are shown along with the ultimate equilibrium. Clearly, our algorithm handles this case perfectly well, providing some encouraging evidence of its wider applicability.

\subsection{Lenard-Balescu equation} \label{sec:num_QLB}
Relaxation problems in which the electrons are much hotter than the ions can exhibit the coupled mode effect in which an ion acoustic mode is excited, leading to a lowering of the energy transfer rate between electrons and ions. This effect has been discussed in many places, from both theoretical \cite{Dharma-Wardana,Vorberger,Benedict17,Benedict} and computational \cite{Chapman,Scullard18} perspectives. Near the frequency of an ion acoustic mode, the dielectric function contributes a sharp peak to the integrand in Eq. (\ref{eq:CnlkXY}) which can be a challenge to capture accurately \cite{Benedict17,Chapman} in rate calculations, let alone quickly enough to be used in a kinetic solver. In Ref. \cite{Scullard18}, it was shown that if one performs the integration over the wavenumber ($X$ in our case) first, the remaining integral over the frequency ($Y$) can be accurately handled with Gaussian quadrature. This is exactly the approach we use here. At $t=0$, when the distributions are Maxwellian and the expansions use the true values of $\beta$, the equilibration rate, i.e., the derivative of the ion temperature, is given by
\begin{equation}
 \frac{d T_i}{dt}=-\frac{C^{(ie)}_{100}}{\beta_i}. \label{eq:dTidt}
\end{equation}
This rate has been thoroughly studied with the quantum Lenard-Balescu equation, with theories including the effects of quantum degeneracy \cite{Vorberger, Chapman, Scullard18, Daligault} and strong coupling \cite{Daligault}. In Table \ref{table:CM}, we compare the rate calculated with Eq. (\ref{eq:dTidt}) with the computations done in Ref. \cite{Scullard18} in which both a careful numerical integration over the double integral and an analytic method similar to ours were used. Although that work included the effects arising from Fermi statistics, the conditions we are considering here are sufficiently non-degenerate that this should have no impact on the answer. Indeed, as can be seen in the table, our computation exactly agrees with that of Ref. \cite{Scullard18}. 

We show in the fourth column of Table \ref{table:CM} the effect of setting $B_{nn'n''}=0$; the change is negligible, and is similar to what one finds by neglecting Eq. (64) of Ref. \cite{Scullard18}. That the $B$ coefficients do not change the answer significantly is rather convenient; their computation is the most expensive part of the reprojections. However, we cannot rule out the possibility that they may be more important for some problems.

\begin{figure}
 \begin{center}
  \includegraphics[width=3.5in]{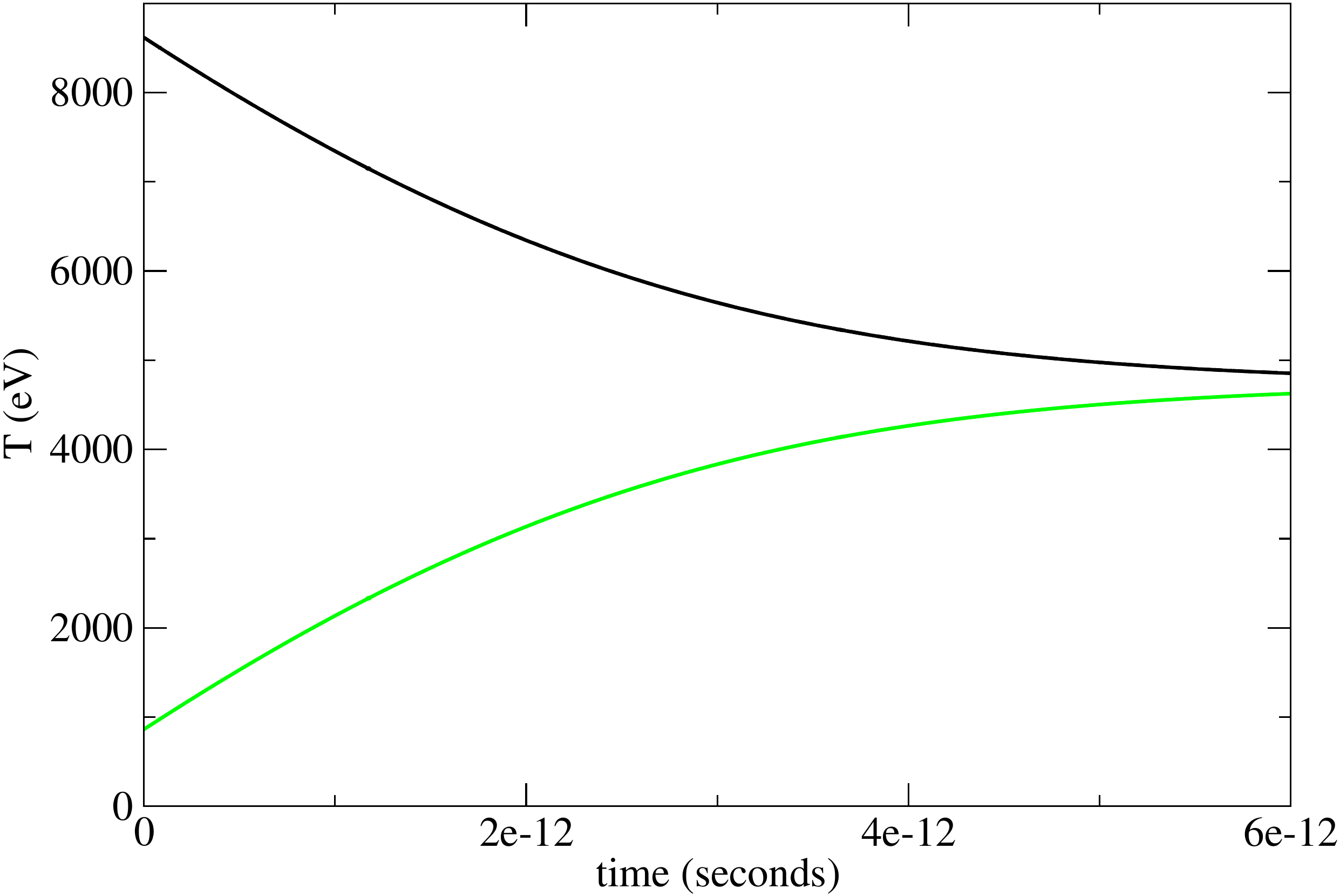}
 \caption{Electron (black) and proton (green) temperatures under the Landau and quantum Lenard-Balescu equations for $n_e=n_i=10^{25}\ \mathrm{cm}^{-3}$, $T_e= 10^8 K$ and $T_i=10^7 K$. The differences between the two equations are not visible in the plot.}
 \label{fig:Te1e8K_Ti1e7K_n1e25}
 \end{center}
\end{figure}

\begin{figure}
 \begin{center}
  \includegraphics[width=3.5in]{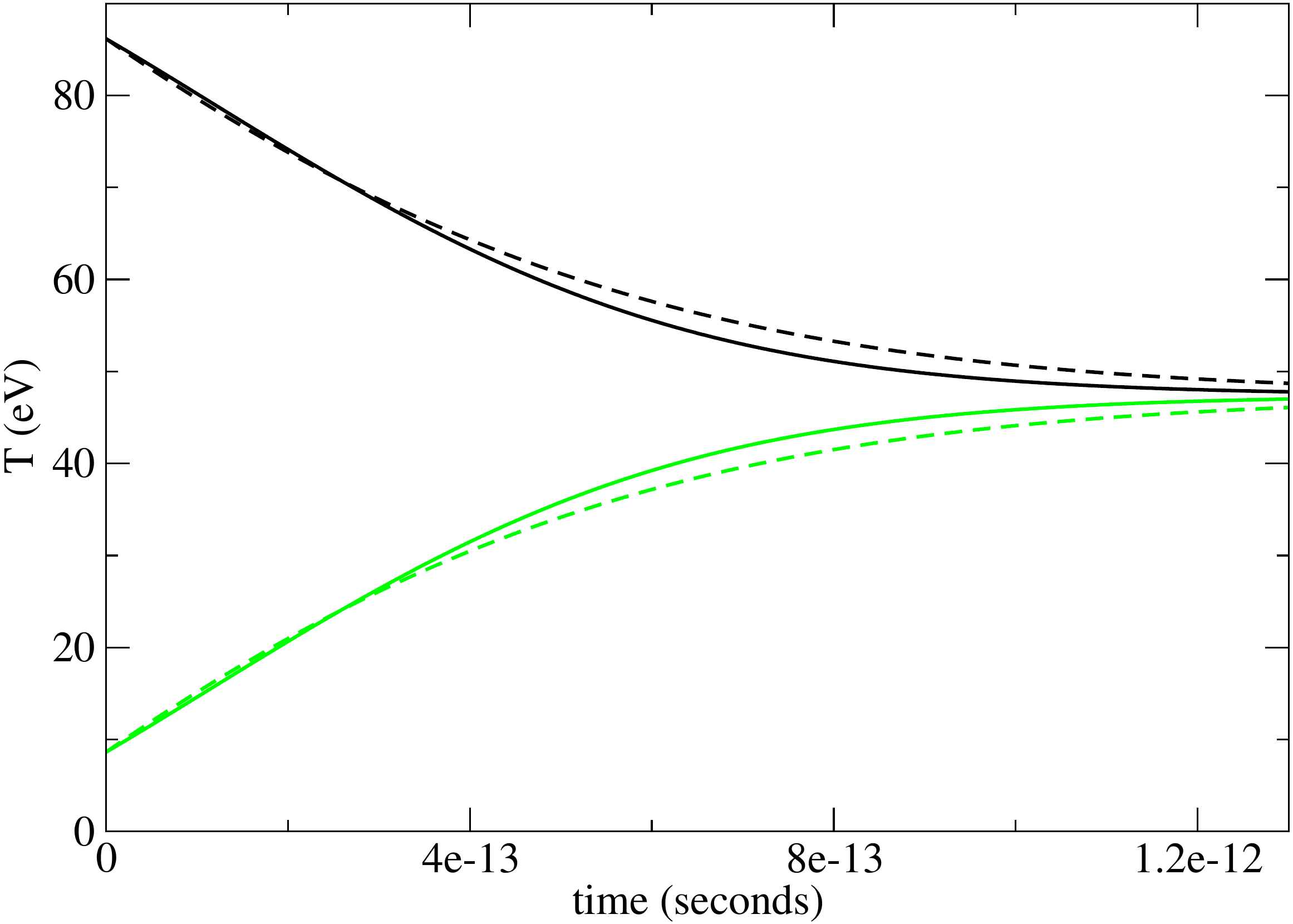}
 \caption{Electron (black) and proton (green) temperatures under the Landau (solid) and quantum Lenard-Balescu (dashed) equations for $n_e=n_i=10^{25}\ \mathrm{cm}^{-3}$, $T_e= 10^6 K$ and $T_i=10^5 K$. The differences between the two equations are visible but minor.}
 \label{fig:Te1e6K_Ti1e5K_n1e23}
 \end{center}
\end{figure}

\begin{figure}
 \begin{center}
  \includegraphics[width=3.5in]{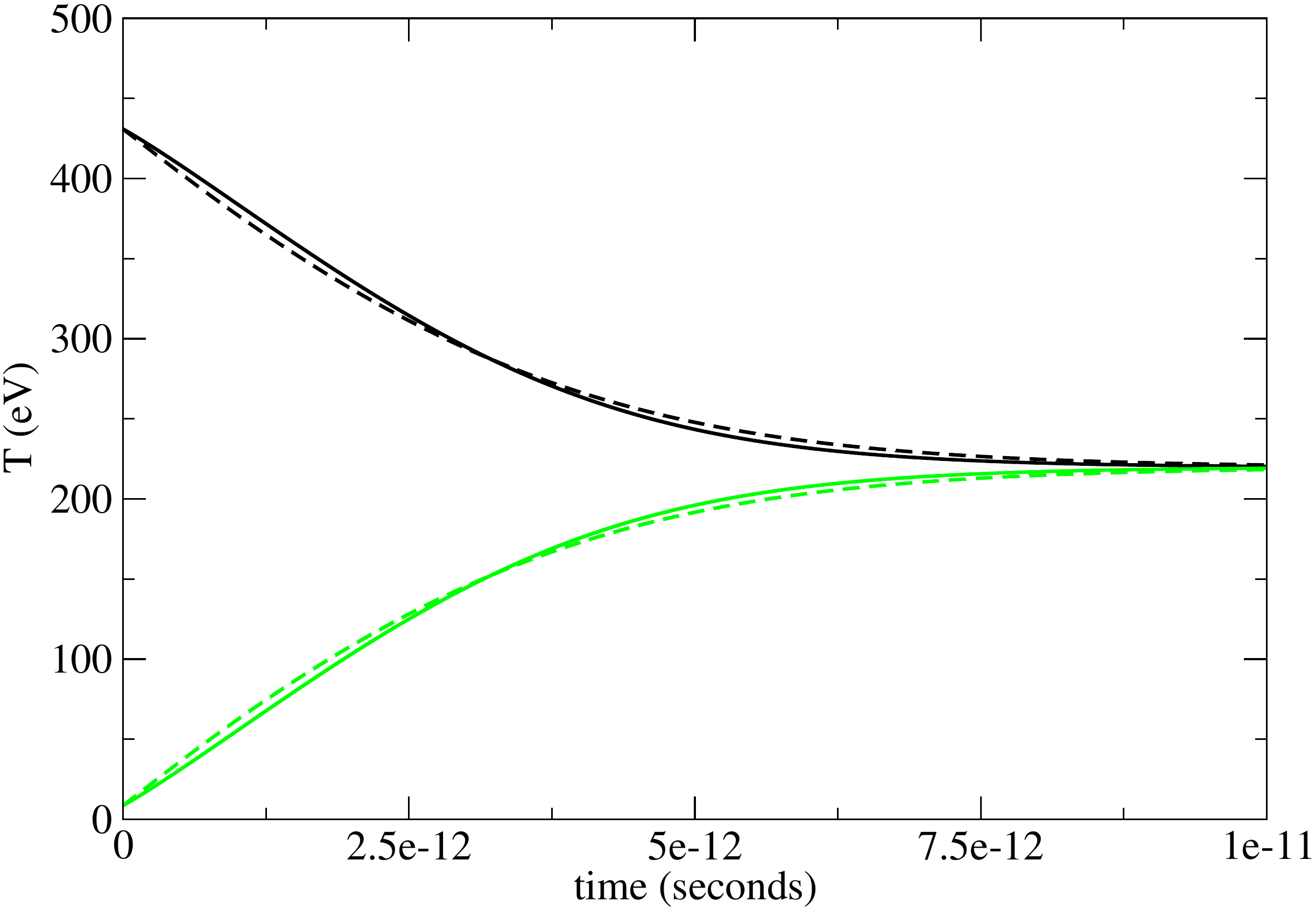}
 \caption{Electron (black) and proton (green) temperatures under the Landau (solid) and quantum Lenard-Balescu (dashed) equations for $n_e=n_i=10^{25}\ \mathrm{cm}^{-3}$, $T_e= 5 \times 10^6 K$ and $T_i=10^5 K$. The differences between the two equations are visible but minor, despite temperatures being separated by a factor of 50.}
 \label{fig:Te5e6K_Ti1e5K_n1e23}
 \end{center}
\end{figure}

Finally, in Figures \ref{fig:Te1e8K_Ti1e7K_n1e25} to \ref{fig:Te5e6K_Ti1e5K_n1e23}, we compare temperature evolutions computed with the QLB equation and the Landau equation. For the Coulomb logarithm, we choose a $\Lambda$ that varies with time to match the logarthmic term we calculated for the Lenard-Balescu equation. That is, we use equation (\ref{eq:Ccalc}) with $B_{nn'n''}=F_{nn'n''}=0$ and we recompute $\eta$ from (\ref{eq:eta}) at every time step based on the current conditions. In Figure \ref{fig:Te1e8K_Ti1e7K_n1e25} we show the temperature evolution for hydrogen with $n_i=n_e=10^{25}\ \mathrm{cm}^{-3}$ and initial temperatures $T_e=10^8 K$ and $T_i=10^7K$. The electrons are shown in black and the protons in green. Both the Landau and Lenard-Balescu solutions are plotted but there are completely indistinguishable from each other. Lowering the density to $n_i=n_e=10^{23}\ \mathrm{cm}^{-3}$, we show the temperature evolution for $T_i = 10^5K$ and $T_e = 10^6K$ in Figure \ref{fig:Te1e6K_Ti1e5K_n1e23}, and $T_e=5 \times 10^6K$ in Figure \ref{fig:Te5e6K_Ti1e5K_n1e23}. In this case, there are visible differences between the Landau and quantum Lenard-Balescu equations but, even when the temperatures are separated by a factor of $50$, these differences are very small. It is possible that by going to even more unrealistically large temperature separations, like those listed in Table \ref{table:CM}, one might see more pronounced differences between the solutions. However, it becomes difficult to integrate the ordinary differential equations for the Lenard-Balescu equation at these conditions. The reason for this is unclear. It could perhaps be that we are beginning to violate the time scale assumptions underlying the derivation of the equation \cite{IchimaruBook} and solutions are becoming unphysical. Perhaps there is a more mundane explanation, but as these conditions are highly unrealistic physically we do not pursue the issue here.

\begin{table*}
\begin{center}
 \begin{tabular}{|c|c|c|c|c|c|}
 \hline
 $n_i$ & Eq. (\ref{eq:dTidt}) & Ref. \cite{Scullard18} & Eq. (\ref{eq:dTidt}), $B_{nn'n''}=0$ & Landau \\
 \hline \hline
 $1 \times 10^{23}$  & 0.0297 & 0.0296 & 0.0292 & 0.0200\\
 $1 \times 10^{24}$  & 0.225  & 0.225  & 0.221  & 0.128 \\
 $1 \times 10^{25}$  & 1.54   & 1.54   & 1.49   & 0.568 \\
 \hline
 \end{tabular}
\caption{The rate of change of the ion temperature for fully ionized hydrogen ($n_e=n_i$) with $T_e=3.0 \times 10^7 K$ and $T_i = 1.0 \times 10^5 K$, calculated from Eq. (\ref{eq:dTidt}) and compared with the results in Ref. \cite{Scullard18}. At these conditions, coupled modes are important, and one must capture accurately the details of the dielectric function in the integration. Clearly, our method is doing this. In the fourth column, we show the calculation done when the $B_{nn'n''}$ of Eq. (\ref{eq:Ccalc}) are set to zero. Although some difference can be seen, it is minimal.}
 \label{table:CM}
 \end{center}
\end{table*}

\section{Conclusion}
We have presented an adaptive spectral method for solving kinetic equations, such as the Landau and quantum Lenard-Balescu equations. Our method identically conserves both energy and particles, while the adaptive scheme greatly mitigates the large memory requirements normally attendant on this spectral approach.

We demonstrated the potential of this algorithm by using it to solve the quantum Lenard-Balescu equation, a problem that would be very difficult with standard Fokker-Planck methods \cite{Epperlein,Chang}. We compared the solutions of electron-ion equilibration problems calculated with the Landau and Lenard-Balescu equations and find that, consistent with other studies into these equations, there is little advantage to solving the full Lenard-Balescu equation, at least for these problems. 

This method can be generalized to distributions that are spatially dependent and non-isotropic in velocity, and we hope that it will be used for such problems in the future. It is possible that among the expanded range of problems that will then be accessible, some will provide a good reason to go beyond the Landau equation.

\section*{Acknowledgments}
We thank \"Omer F. Tekin and Hai Le for helpful discussions. Part of this research was performed while the authors were visiting the Institute for Pure and Applied Mathematics (IPAM), which is supported by the National Science Foundation. This work was performed under the auspices of the U.S. Department of Energy at the Lawrence Livermore National Laboratory under Contract No. DE-AC52-07NA27344.

\bibliography{scullard}

\appendix
\section{Reprojection formula} \label{sec:reprojection}
We wish to take a distribution $f(v)$ in the form:
\begin{equation}
    f^{\mathrm{eq}}_{\beta}(v) \sum_{j=0}^{n_{\mathrm{max}}} A_j(\beta) L_j^{\left(\frac{1}{2}\right)}\left(\frac{mv^2\beta}{2}\right)
\end{equation}
where $\{A_n(\beta)\}$ are already calculated, and express it in the form
\begin{equation}
    f^{\mathrm{eq}}_{\beta'}(v) \sum_{j=0}^{\infty} A_j(\beta') L_j^{\left(\frac{1}{2}\right)}\left(\frac{mv^2\beta'}{2}\right);
\end{equation}
so that 
\begin{equation}
      f^{\mathrm{eq}}_{\beta}(v) \sum_{j=0}^{n_{\mathrm{max}}} A_j(\beta) L_j^{\left(\frac{1}{2}\right)}\left(\frac{mv^2\beta}{2}\right) =  f^{\mathrm{eq}}_{\beta'}(v)\sum_{j=0}^{\infty} A_j(\beta') L_j^{\left(\frac{1}{2}\right)}\left(\frac{mv^2\beta'}{2}\right).
\end{equation}

We begin evaluating this by multiplying on both sides by $L_n^{(1/2)}\left(m\beta'v^2/2 \right)$ and integrating both sides by $\int d^3v$. Our equation becomes
\begin{eqnarray}
  & & \int_{0}^\infty dv f_\beta^{\mathrm{eq}}(v) \sum_{j=0}^{n_{\mathrm{max}}} A_j(\beta) L_j^{\left(\frac{1}{2}\right)}\left(\frac{m\beta v^2}{2}\right) L_n^{\left(\frac{1}{2}\right)}\left(\frac{m\beta'v^2}{2}\right) v^2 \cr 
    &=& \int_{0}^\infty dv f_{\beta'}^{\mathrm{eq}}(v) \sum_{j=0}^{\infty} A_j(\beta') L_j^{\left(\frac{1}{2}\right)}\left(\frac{m\beta'v^2}{2}\right) L_n^{\left(\frac{1}{2}\right)}\left(\frac{m\beta'v^2}{2}\right) v^2. 
\end{eqnarray}

Evaluating the right hand side gives us
\begin{eqnarray}
& &\int_{0}^\infty dv \hspace{1 mm} v^2 \left(\frac{m\beta}{2\pi}\right)^{3/2} e^{-\frac{m \beta v^2}{2}}  \sum_{j=0}^{\infty} A_j(\beta)   L_j^{\left(\frac{1}{2}\right)}\left(\frac{m\beta v^2}{2}\right) L_n^{\left(\frac{1}{2}\right)}\left(\frac{m\beta'v^2}{2}\right) \cr
&=& \frac{1}{2} \left(\frac{1}{\pi}\right)^{3/2} A_n(\beta')\frac{\Gamma(n+3/2)}{n!}.
\end{eqnarray}
We simplify this by changing the bounds of the integral (using the fact that our function is even), and by using the substitution $x^2 \equiv m\beta v^2/2$. Our equation becomes: 

\begin{eqnarray}\label{D-27}
 & &  \frac{1}{2} \left(\frac{1}{\pi}\right)^{3/2} \sum_{j=0}^{n_{\mathrm{max}}} A_j(\beta) \int_{-\infty}^\infty dx x^2 e^{-x^2} L_j^{\left(\frac{1}{2}\right)}(x^2) L_n^{\left(\frac{1}{2}\right)}\left(\frac{\beta'}{\beta} x^2\right) \cr
 &=& \frac{1}{2} \left(\frac{1}{\pi}\right)^{3/2} A_n(\beta')\frac{\Gamma(n+3/2)}{n!}.
\end{eqnarray}

We can now replace our Laguerre polynomials with Hermite polynomials using 
\begin{equation}
    L_{j}^{\left(\frac{1}{2}\right)}(x^2)= \frac{H_{2n+1}(x)}{x(-1)^j 2^{2j+1} j!}
\end{equation}

\begin{equation}
    L_n^{\left(\frac{1}{2}\right)}\left(\frac{\beta'}{\beta}x^2\right) = \frac{H_{2n+1}\left(x\sqrt{\frac{\beta'}{\beta}}\right)}{x\sqrt{\frac{\beta'}{\beta}}(-1)^{n}2^{2n+1}n!}
\end{equation}.

This turns the left hand side of (\ref{D-27}) into

\begin{eqnarray} \label{D-28}
& & \frac{1}{2} \left(\frac{1}{\pi}\right)^{3/2}\sqrt{\frac{\beta}{\beta'}}  \sum_{j=0}^{n_{\mathrm{max}}} A_j(\beta)\frac{1}{(-1)^{j+n} 2^{2n+2j+2}j! n!}\cr
& & \ \ \ \ \ \ \ \ \ \ \ \ \ \times \int_{-\infty}^\infty dx e^{-x^2} H_{2j+1}(x) H_{2n+1}\left( x \sqrt{\frac{\beta'}{\beta}}\right). 
\end{eqnarray}
When $j>n$ our integral is equal to zero. Otherwise (\ref{D-28}) becomes
\begin{equation}
    \frac{1}{\pi}\sum_{j=0}^{n} A_j(\beta) \frac{\left(\frac{\beta'}{\beta}\right)^j \left(\frac{\beta'}{\beta} -1\right)^{n-j}(2n+1)!}{4^{n+1} (-1)^{n+j}j! n! (n-j)!}.
\end{equation}
Now, using the identity
\begin{equation}
\frac{\sqrt{\pi}(2n+1)!}{2^{2n+1}\Gamma\left(n+3/2\right)}=n!,
\end{equation}
we find (\ref{eq:Aprime}).

\section{Calculating the energy} \label{sec:betaprime}
In order to do the reprojection, we need to know the value of $\beta'$ at time $t$ in terms of the expansion parameter $\beta$. Assume that at time $t$ we know the coefficients ${A_i}$ for the Laguerre expansion of $f$ in terms of $\beta$. The kinetic energy ($\mathrm{KE}$) of a species is given by 
\begin{eqnarray}\label{D-31}
  \mathrm{KE}&=&\frac{1}{2} m \int d^3 \mathbf{v} f_i(\mathbf{v}, t) v^2
    = \frac{1}{2} m \int d^3 \mathbf{v} f_\beta^{\mathrm{eq}}(v) \sum_{j=0}^{n_{\mathrm{max}}} A_j(\beta) L_j^{\left(\frac{1}{2}\right)}\left(\frac{m\beta v^2}{2}\right) v^2 \cr
    &=& \frac{1}{2}m \sum_{j=0}^{n_{\mathrm{max}}
    } A_j(\beta) 4\pi\int_{0}^\infty dv \hspace{1 mm} v^4 n_i \left(\frac{m\beta}{2\pi}\right)^{3/2}e^{\frac{-m\beta v^2}{2}}L_j^{\left(\frac{1}{2}\right)}\left(\frac{m\beta v^2}{2}\right) \cr
    &=& \sum_{j=0}^{n_{\mathrm{max}}} \frac{2n_i}{\beta\sqrt{\pi}}A_j(\beta)\int_{0}^\infty dx \hspace{1 mm} x^{3/2}e^{-x}L_j^{\left(\frac{1}{2}\right)}(x) \cr
    &=& \sum_{j=0}^{n_{\mathrm{max}}} \frac{2n_i}{\beta\sqrt{\pi}}A_j(\beta)\left(\int_{0}^\infty dx \hspace{1 mm} x^{1/2}e^{-x}L_j^{\left(\frac{1}{2}\right)}(x)\left(x-\frac{3}{2}\right) \right. \cr
  & & \ \ \ \ \ \ \ \ \ \ \ \ \  \left. +\frac{3}{2} \int_{0}^\infty dx \hspace{1 mm} x^{1/2}e^{-x}L_j^{\left(\frac{1}{2}\right)}(x) \right).
\end{eqnarray}
where $n_i$ is the number density of the species under consideration. Using the fact that $L_0^{\left(\frac{1}{2}\right)}(x) = 1$ and $L_1^{\left(\frac{1}{2}\right)}(x) = \frac{3}{2} - x$, (\ref{D-31}) becomes
\begin{equation}
  \mathrm{KE} =\frac{2n_i}{\beta\sqrt{\pi}}[-\Gamma{(5/2)}A_1(\beta) + \frac{3}{2}\Gamma{(3/2)}A_0(\beta)].
\end{equation}
Noting that $A_0(\beta) = 1$ and $\Gamma{(5/2)} = (3/2)\Gamma{(3/2)} = \frac{3}{4}\sqrt{\pi}$ this simplifies to
\begin{equation}
 \mathrm{KE} = \frac{3n_i}{2\beta}\left(1 - A_1(\beta)\right). \label{C-5}
\end{equation}
The energy of a particle in a Maxwell distribution at $\beta'$ is given by 
\begin{equation}
 \mathrm{KE} = \frac{3 n_i}{2 \beta'}, \label{C-6}
\end{equation}
and setting Eq. (\ref{C-5}) equal to Eq. (\ref{C-6}) we get Eq. (\ref{eq:betaprime}).

\end{document}